# Origin of the Ocean on the Earth: Early Evolution of Water D/H in a Hydrogen-rich Atmosphere

## Hidenori Genda[1] and Masahiro Ikoma[2]

[1]*Research Center for the Evolving Earth and Planets, Tokyo Institute of Technology,
2-12-1 Ookayama, Meguro-ku, Tokyo 152-8551, Japan*
[2]*Department of Earth and Planetary Sciences, Tokyo Institute of Technology,
2-12-1 Ookayama, Meguro-ku, Tokyo 152-8551, Japan*

Phone:   +81-3-5734-3936
Fax:     +81-3-5734-3538
E-mail:  genda@geo.titech.ac.jp    (Genda)
         mikoma@geo.titech.ac.jp   (Ikoma)

Pages:   29
Figures:  6

ABSTRACT

The origin of the Earth's ocean has been discussed on the basis of deuterium/hydrogen ratios (D/H) of several sources of water in the solar system. The average D/H of carbonaceous chondrites (CC's) is known to be close to the current D/H of the Earth's ocean, while those of comets and the solar nebula are larger by about a factor of two and smaller by about a factor of seven, respectively, than that of the Earth's ocean. Thus, the main source of the Earth's ocean has been thought to be CC's or adequate mixing of comets and the solar nebula. However, those conclusions are correct only if D/H of water on the Earth has remained unchanged for the past 4.5 Gyr. In this paper, we investigate evolution of D/H in the ocean in the case that the early Earth had a hydrogen-rich atmosphere, the existence of which is predicted by recent theories of planet formation no matter whether the nebula remains or not. Then we show that D/H in the ocean increases by a factor of 2–9, which is caused by the mass fractionation during atmospheric hydrogen loss, followed by deuterium exchange between hydrogen gas and water vapor during ocean formation. This result suggests that the apparent similarity in D/H of water between CC's and the current Earth's ocean does not necessarily support the CC's origin of water and that the apparent discrepancy in D/H is not a good reason for excluding the nebular origin of water.

# 1. Introduction

Although the mass of the ocean on the Earth (~ $1.4 \times 10^{21}$ kg) is a tiny fraction (~ 0.02 %) of the bulk Earth mass, the existence of the ocean is sufficient to distinguish the Earth from the other planets in the solar system. In particular, the existence of sufficient liquid water is thought to be essential for the origin and evolution of life.

There are two fundamental questions about the origin of the ocean on the Earth: When did the ocean form, and where did the water come from? With regard to the former question, we have a piece of geological evidence that constrains the age of the ocean. Isua Supracrustal rocks in West Greenland (Rb-Sr age of 3.8 Gyr) include metavolcanics and metasediments (e.g., Appel *et al.* 1998). The existence of sediments implies that a substantial amount of liquid water (i.e., ocean) already existed on the Earth at least 3.8 Gyr ago. While there is no direct answer to the latter question, the ratios of deuterium to hydrogen in the present Earth's ocean and some possible water sources have often been discussed.

Three possible sources of water on the Earth have been proposed so far; water-containing rocky planetesimals like carbonaceous chondrites (CC's), icy planetesimals like comets, and the solar nebula. If water-containing planetesimals accrete to form the Earth, impact-induced degassing of water from the planetesimals forms a massive steam atmosphere that ends up the ocean after its cooling (Lange and Ahrens 1982; Matsui and Abe 1986; Zahnle *et al.* 1988). Water-containing rocky planetesimals (e.g., Morbidelli *et al.* 2000; Raymond *et al.* 2004, 2005) and/or icy planetesimals (e.g., Gomes *et al.* 2005) can be delivered from outer regions (typically > 2–3 AU) because of gravitational perturbation of the giant planets. If the solar nebula survives after the completion of the Earth, a massive hydrogen-rich atmosphere is formed on the Earth. The atmospheric hydrogen reacts with oxides such as FeO contained in the magma ocean of the Earth to produce a sufficient amount of water (Sasaki 1990; Ikoma and Genda 2006). However, because of several uncertainties of planet formation processes, the origin of the ocean on the Earth is still a matter of controversy.

Figure 1 shows the number ratios of deuterium to hydrogen (D/H) in various possible sources of the ocean on the Earth. The average D/H ratio in carbonaceous chondrites is very close to that in the present seawater. The D/H ratios in the three comets and the nebular hydrogen are higher by a factor of 2 and smaller by a factor of ~7, respectively, than that of the present seawater. Based on these data, seawater on the Earth has been generally believed to have come from carbonaceous chondrites. Another possibility is the adequate mixing of water from comets and the solar nebula that yields the D/H ratio of the present seawater.

The above discussion assumes that the D/H ratio of water on the Earth has remained unchanged for the past 4.5 Gyr. However, the assumption must be ascertained, because the D/H ratio of water, in principle, changes during the formation and evolution of the ocean. As shown later in this paper, the D/H ratio would have undergone appreciable changes, especially if a massive hydrogen-rich atmosphere (comparable in mass with water) had been formed on the early Earth. The existence of a hydrogen-rich atmosphere on the early Earth is suggested by recent theories of terrestrial planet formation.

Recent theories about terrestrial planet formation in the solar system suggest that the nebular gas remained until the terrestrial planets were completed (Kominami and Ida 2002; Nagasawa *et al*. 2005). The reason why the persistence of the nebular gas is favored is that it can account for the current low eccentricities of the terrestrial planets. Without damping processes of the planets' eccentricities, fully-formed terrestrial planets are known to have higher eccentricities than the current values (e.g., Chambers *et al*. 1996). Although, as demonstrated by Kominami and Ida (2002), the required nebular density is as low as $10^{-4}$–$10^{-3}$ times that of the minimum-mass solar nebula (Hayashi 1981), the value of nebular density is high enough for an Earth-mass planet to have a massive hydrogen-rich atmosphere of more than $10^{21}$ kg (Ikoma and Genda 2006).

There are other theories (e.g. Raymond *et al*. 2005), in which no nebular gas is needed to produce the low eccentricities of the terrestrial planets. Because of gravitational perturbation by Jupiter, water-containing chondritic planetesimals come to the terrestrial planet region from the region corresponding to the asteroid belt. In the context of those theories, dynamical friction due to the planetesimals is effective in lowering the eccentricities of fully-formed planets. Even in this scenario, a hydrogen-rich atmosphere is likely to be formed for the following reasons. If the planetesimals contain Fe metal, reduction of water to hydrogen by Fe metal in magma ponds or magma ocean produces a hydrogen-rich atmosphere; in that case, the molar $H_2/H_2O$ ratio is ~ 10, if the oxygen fugacity is buffered, for example, by FeO (Kuramoto and Matsui 1996). Even if the planetesimals contain no Fe-metal (e.g., carbonaceous chondrites), recent calculations of chemical-equilibrium composition of gas from carbonaceous chondrites show that reducing components such as $H_2$ and $CH_4$ are produced (Hashimoto *et al*. 2007). This is because carbonaceous chondrites contain more hydrogen and carbon (in the form of organics) than trivalent iron oxide such as $Fe_2O_3$, so that hydrogen and carbon are not fully oxidized by $Fe_2O_3$.

In this paper, we examine the evolution of the D/H ratio of water on the Earth in the case where the early Earth had a hydrogen-rich atmosphere. The surface environment at the time of the Earth's formation is hot enough for water above the surface to be completely in the vapor phase. As the atmosphere cools, the water vapor condenses to form an ocean. After that, the atmosphere escapes to the space because of intense solar UV irradiation. We thus focus on three processes: deuterium exchange between hydrogen gas and water vapor, that between hydrogen gas and liquid water (i.e., the ocean), and mass fractionation during hydrodynamic escape of the atmosphere. In Section 2, we investigate the isotopic equilibrium between hydrogen gas and water vapor. In Sections 3, we discuss the timescales on which several processes relevant to the evolution of the D/H ratio occur. In Section 4, we calculate the D/H ratio in the ocean, neglecting deuterium exchange between the ocean and the atmosphere and mass fractionation during atmospheric escape, in order to constrain the lower limit to the enhancement of the D/H ratio. In Section 5, including those effects, we simulate long-term evolution of the D/H ratio in the ocean. In Section 6 we discuss the origin of the Earth's ocean, from the viewpoint of the D/H ratio in the ocean.

## 2. Isotopic Equilibrium and Reaction Kinetics

We first consider deuterium exchange between hydrogen gas and water vapor. The deuterium exchange occurs via the following reaction (e.g., Robert *et al*. 2000),

$$HD + H_2O \rightleftarrows HDO + H_2. \tag{1}$$

We ignore $D_2$ and $D_2O$ molecules, because the D/H ratio considered in this paper is so small (~$10^{-4}$–$10^{-5}$) that their contribution to the D/H ratio is negligible.

The equilibrium constant $K$ for reaction (1) is given by

$$K = \frac{p_{HDO}\, p_{H_2}}{p_{HD}\, p_{H_2O}}, \tag{2}$$

where $p_X$ is the partial pressure of species X. The D/H ratios of hydrogen gas and water vapor are respectively defined as

$$(D/H)_h = \frac{p_{HD}}{2p_{H_2} + p_{HD}} \quad \text{and} \quad (D/H)_{w(g)} = \frac{p_{HDO}}{2p_{H_2O} + p_{HDO}}. \tag{3}$$

Since $p_{H2} \gg p_{HD}$ and $p_{H2O} \gg p_{HDO}$ in the situation considered in this paper, $K$ is approximately given by

$$K \cong \frac{(D/H)_{w(g)}}{(D/H)_h}. \tag{4}$$

Based on the experiments for 273–1000K performed by Richet et al. (1997), we use the following approximate formula for $K$;

$$K = \frac{0.22 \times 10^6}{T^2} + 1, \tag{5}$$

where $T$ is the temperature in Kelvin. Note that pressure dependence of $K$ is negligible (Richet *et al*. 1977). We have drawn a graph of the $K$ in Fig. 2a. The values of $K$ are found to be higher than 1, which means that water vapor is enriched in deuterium relative to hydrogen gas: The enrichment is significant for low temperatures. Although massive hydrogen-rich atmospheres of interest in this paper would have high effective vibrational temperature $T_v$ (Parkinson *et al.* 1999; Parkinson *et al.* 2006), changes in $K$ due to high $T_v$ are as small as less than 1% for $T_v$ = 273–1273 K (Richet *et al.* 1977).

For later discussion, we estimate the characteristic reaction time $\tau_e$ in which isotopic equilibrium between hydrogen gas and water vapor is attained. We define $\tau_e$ by

$$\tau_e = -(D/H)^d_{w(g)} \bigg/ \frac{d}{dt}\left[(D/H)^d_{w(g)}\right], \qquad (6)$$

where $(D/H)^d_{w(g)}$ is deviation from the equilibrium value of the D/H ratio of water vapor. Since $(D/H)_h \ll 1$ and $(D/H)_{w(g)} \ll 1$ in this paper, $\tau_e$ is expressed by (see Appendix A)

$$\tau_e = \frac{1}{k^+ p_{H_2O} + k^- p_{H_2}}, \qquad (7)$$

where $k^+$ and $k^-$ is the rate constants for the forward and backward reactions in Eq. (1), respectively. Furthermore, because $k^+ = K \times k^-$ and $K$ is of the order of 1 for temperatures of interest as found above, Eq. (7) can be written by

$$\tau_e \sim \frac{1}{k^- p_t}, \qquad (8)$$

where $p_t$ is the total pressure ($p_{H2O} + p_{H2}$). For $k^-$, we adopt the widely-used formula,

$$k^- = 2.0 \times 10^{-28} \exp\left[-\frac{5178}{T}\right] \quad m^3 s^{-1}, \qquad (9)$$

which was given by Lécluse and Robert (1994) based on their experiments. Figure 2b shows $\tau_e$ against $T$ for two different values of $p_t$, which is calculated from Eqs. (8) and (9). As a matter of course, higher $T$ or $p_t$ yields shorter $\tau_e$.

Note that the rate constant for reaction (1) we have adopted is much higher than values theoretically predicted from a suite of elemental reactions whose rate constants are well known and those predicted from values usually reported in the literature on this kind of isotopic exchange reactions (e.g., $H_2 + D_2 \rightarrow HD + HD$). Such prediction is, however, difficult because isotopic exchange reactions often involve several unrecognized pathways. Since the exchange fraction experimentally measured by Lécluse and Robert (1994) was much smaller than those usually reported, and since this fraction was close to the observed natural variations of the D/H ratio (Robert 2006, personal communication), we have used the rate constant published by Lécluse and Robert (1994). There is no experimental data of $k^-$ other than theirs.

## 3. Relevant Processes and Their Typical Timescales

If the Earth initially had a massive atmosphere composed of hydrogen and water, the proportion of deuterium in water increases through the following processes. When the atmosphere is hot enough for all the water to be in the vapor form, the gas-phase reaction for deuterium exchange makes water enriched in deuterium, as described in section 2. Then, as the atmosphere cools down, water vapor condenses and falls to form an ocean. During and/or after the ocean formation, there occurs deuterium exchange between the ocean and the hydrogen-rich atmosphere through circulation of water in the atmosphere-ocean system. Furthermore, such a large amount of hydrogen must have been lost almost completely, because the present atmosphere contains only a tiny amount of hydrogen. Mass fractionation due to the atmospheric escape results in deuterium enrichment in the atmosphere and also the ocean. Before examining the evolution of the D/H ratio of water on the Earth, we discuss the timescales of those processes.

The ocean formation is controlled by cooling of the atmosphere with saturated water vapor. In the case of a $H_2O-CO_2$ atmosphere with $H_2O$ of 100–1000 bar, cooling is known to occur on timescale of $10^2-10^3$ years (Abe 1993). The cooling timescale for a $H_2O-H_2$ atmosphere would be also of the same order, because the radiative flux from the atmosphere with saturated water vapor is mainly determined by optical and thermodynamic properties of $H_2O$ and the dominant source of the atmospheric heat content is the latent heat of $H_2O$ (Abe 1993). Compared to this cooling timescale, the reaction timescale $\tau_e$ is much shorter (e.g., $\tau_e < \sim 0.1$ years for $\sim 1000$ bar and $> 300$ K; see Fig. 2b). This means that the D/H ratio of water vapor always achieves its equilibrium value during the cooling. Thus, the liquid water that have formed at a given temperature would have the equilibrium value of the D/H ratio of water vapor at that temperature.

Deuterium exchange between the liquid water (i.e., the ocean) and the hydrogen gas (i.e., the atmosphere) would also occur through circulation of water in the atmosphere-ocean system during and after the ocean formation. The timescale on which the entire ocean isotopically equilibrates with the atmosphere ($\tau_{AO}$) can be estimated as

$$\tau_{AO} \approx \frac{\tau_e}{\tau_s} \times \tau_c, \qquad (10)$$

where $\tau_s$ is the typical time for which the water vapor that has evaporated from the ocean stays in the atmosphere and $\tau_c$ is the typical time in which all the water in the ocean vaporizes and circulates through the atmosphere. $\tau_e/\tau_s$ corresponds to the number of the circulations required for the water to isotopically equilibrate with the atmosphere. On the present Earth, $\tau_s \sim 10$ days. The value of $\tau_s$ for the early Earth with a hydrogen-rich atmosphere is uncertain, but probably larger. This is because the pressure scale height of such an atmosphere is larger compared to that of the current atmosphere because of small molecular weight, which means that the altitude of the tropopause would be higher, and thus water vapor stays longer in the atmosphere. Given that $\tau_c \sim d_o/d_r$ ($d_r$ being the annual rainfall and $d_o$ being the mean ocean depth), $\tau_c$ is estimated at

~3000 years for the current Earth because $d_r \sim 1$ m/yr and $d_o = 3$ km. Since it is generally thought that the Earth have been as warm as present over 4.5Gyr, $\tau_c$ is probably similar to the current value for the early Earth. In any case, deuterium exchange between the ocean and the atmosphere takes place rather fast; if there exists a hydrogen-rich atmosphere that is comparable in mass to the present Earth's ocean (i.e., $p_{H2} \sim 3000$ bar and $T \sim 300$ K, yielding $\tau_e < 0.1$ years; see Fig. 2a), $\tau_{AO}$ is estimated at as short as $< \sim 10^4$ years, which is not so different from the timescale of ocean formation.

Finally, atmospheric escape due to intense stellar XUV (X-ray and extreme-UV radiation with wavelength of $<\sim 100$ nm) irradiation occurs on timescale of $10^8$–$10^9$ years (e.g., Sekiya *et al*. 1980; see section 5). This timescale is much longer than that of ocean formation. The escape timescale is also much longer than $\tau_{AO}$. This suggests that the ocean always equilibrates isotopically with the atmosphere during the atmospheric escape.

### 4. Minimum Enhancement of D/H of Water in the Ocean

All the processes described in the previous section enhance the D/H ratio of water. The way to minimize the deuterium enrichment in the ocean is to quench deuterium exchange between the ocean and the atmosphere once the ocean has formed. We have found in the previous section that such a situation is unrealistic because deuterium exchange between the ocean and the atmosphere occur efficiently through circulation of water in the ocean-atmosphere system after (and possibly during) ocean formation. However, in order to discuss the fact that the D/H ratio of water in carbonaceous chondrites are quite similar to that of the current ocean on the Earth (see section 6), it is worthwhile calculating the lower limits to deuterium enrichment in the ocean, before modeling the realistic evolution of the D/H ratio in the ocean.

We consider a box with volume $V$ and temperature $T$ containing hydrogen and water, and make the following assumptions. (1) The gas is ideal, and the volume of liquid water is negligible in the Clausius-Clapeyron relationship that gives the saturation water vapor pressure $p_{w(g)}$ in the form of

$$p_{w(g)}(T) = p^* \exp\left(-\frac{l}{RT}\right), \tag{11}$$

where $p^*$ is the constant for the water vapor saturation curve, $l$ is the latent heat of water, and $R$ is the gas constant ( = 8.314 J mol$^{-1}$ K$^{-1}$). For $p^*$ and $l$ we adopt $1.4 \times 10^{11}$ Pa and $4.37 \times 10^4$ J mol$^{-1}$, respectively (Eisenberg and Kauzmann 1969). (2) Isotopic equilibrium between hydrogen gas and water vapor is always achieved (see section 3). (3) The D/H ratio of the liquid water formed at a given temperature is equal to that of water vapor at that temperature and undergoes no subsequent changes. This means that only addition of newly condensed water molecules changes the D/H ratio of bulk liquid water. Note that the D/H ratio of liquid water is different from that of the surrounding water vapor, because condensabilities of $H_2O$ and HDO differ. However, in the temperature range ($T > 400K$) in which almost all water vapor condenses (see below), the difference is small (< 2%; Friedmann and O'Neil 1977) compared to changes in the D/H ratio considered in this paper. (4) We neglect vertical distribution of water in the atmosphere and temperature changes due to water condensation. These simplifications are reasonable, because the atmospheric mass is concentrated near the planetary surface and because most of the water vapor condenses in a limited range of temperature (~ 500 K; see below).

We introduce a new quantity called the deuterium enrichment factor;

$$f_X = \frac{(D/H)_X}{(D/H)_0}, \qquad (12)$$

where $(D/H)_0$ is the D/H ratio in the entire system, and $(D/H)_X$ is the D/H ratio of species X, namely, hydrogen gas $(D/H)_h$, water vapor $(D/H)_{w(g)}$, and liquid water $(D/H)_{w(l)}$. Under the above assumptions, the changes in D/H ratios of hydrogen gas, water vapor, and liquid water are described by the following set of equations (see Appendix B):

$$f_{w(g)} = K f_h, \qquad (13)$$

$$\frac{df_{w(l)}}{dT} = \left(f_{w(g)} - f_{w(l)}\right) \frac{1}{M_{w(l)}} \frac{dM_{w(l)}}{dT}, \qquad (14)$$

$$\rho_h + \frac{\rho_{w(g)}(T_0)}{9} = f_h \rho_h + \frac{f_{w(l)}}{9}\left\{\rho_{w(g)}(T_0) - \rho_{w(g)}(T)\right\} + \frac{f_{w(g)}}{9} \rho_{w(g)}(T), \qquad (15)$$

$$M_{w(l)} = V\left\{\rho_{w(g)}(T_0) - \rho_{w(g)}(T)\right\}, \qquad (16)$$

where $M_{w(l)}$ is the total mass of liquid water, $\rho_h$ and $\rho_{w(g)}$ are the partial density of hydrogen gas and water vapor (see Eq. (B1) for the definition), and $T_0$ is the temperature at which the water vapor starts to condense. While $\rho_{w(g)}$ and $\rho_{w(l)}$ change with temperature, $\rho_h$ is constant because hydrogen gas does not condense in the temperature range considered here. Equation (13) means the isotopic equilibrium between hydrogen gas and water vapor (assumption (2) and Eq. (4)). Equation (14) represents the change in the D/H ratio of the liquid water by addition of newly condensed water molecules (assumption (3)). Equation (15) is derived from the conservation of the amount of deuterium molecules: the left-hand side of Eq. (15) corresponds to the state at temperature $T_0$, while its right-hand side corresponds to the state at temperature $T$. Equation (16) represents the total mass of liquid water as a function of temperature. Inserting Eq. (16) in Eq. (14), we obtain

$$\frac{df_{w(l)}}{dT} = \left(f_{w(g)} - f_{w(l)}\right) \frac{1}{\rho_{w(g)}(T_0) - \rho_{w(g)}(T)} \frac{d\rho_{w(g)}(T)}{dT}. \tag{17}$$

$\rho_{w(g)}(T)$ is determined by Eq. (11) and equation of state for water vapor. Thus, solving Eqs. (13), (15), and (17), we obtain $f_{w(l)}$, $f_{w(g)}$, and $f_h$ numerically as a function of temperature for given $T_0$ and $\rho_h$.

Figure 3 shows the changes in $f_{w(l)}$, $f_{w(g)}$, and $f_h$ during cooling from $T_0 = 594$ K for $\rho_h = 81$ kg/m$^3$ that corresponds to the vapor pressure $p_{w(g)}(T_0) = 200$ bar and the molar H$_2$/H$_2$O = 10. At $T = T_0$, all the water is in the vapor form and $f_{w(g)} = 1.54$, which is the equilibrium value at $T_0$ determined by $K(T_0)$. As the temperature decreases, the amount of liquid water increases. At $T = 550$K half of the water is liquid ($p_{w(g)} = 99$ bar); at $T = 500$K approximately 80% of the water is liquid ($p_{w(g)} = 38$ bar). Thus, although $f_{w(l)}$ slightly increases below 500K, its final value is almost determined at 500−594 K. In this case shown in Fig. 3, the D/H ratio of liquid water at 300K is higher by a factor of 1.7 (i.e., $f_{w(l)} = 1.7$) than that of the entire system (i.e., [D/H]$_0$).

The final value of $f_{w(l)}$ (e.g., at 300K) depends on $p_{w(g)}(T_0)$ and H$_2$/H$_2$O. Figure 4 shows $f_{w(l)}$ at 300 K[1] for various values of $p_{w(g)}(T_0)$ and H$_2$/H$_2$O. First we find that small $p_{w(g)}(T_0)$ results in large $f_{w(l)}$. Because the amount of water vapor is determined only by $T$ (i.e., saturation-vapor pressure), the amount of liquid water at a given $T$ ($< T_0$) depends on the total amount of water (i.e., $p_{w(g)}(T_0)$). Thus, lower $p_{w(g)}(T_0)$ results in a smaller amount of liquid water at a given $T$ ($< T_0$). This means significant water condensation continues until lower temperatures in the case of smaller $p_{w(g)}(T_0)$. In the case of H$_2$/H$_2$O = 10, for example, 50% of the water is already liquid at 550K if $p_{w(g)}(T_0) = 200$ bar, while only 1% is liquid at the same temperature if $p_{w(g)}(T_0) = 100$ bar. Since the equilibrium value of $f_{w(l)}$ increases as temperature decreases, liquid water becomes more enriched in deuterium when $p_{w(g)}(T_0) = 100$ bar. In Fig. 4, we also find that small H$_2$/H$_2$O results in small $f_{w(l)}$. We can easily understand the result, if we consider the extreme case of no hydrogen molecules (i.e., H$_2$/H$_2$O = 0); $f_{w(l)} = 1$ in that case. If the present Earth's ocean is fully vaporized, the pressure of the steam atmosphere would be ~ 300 bar, because the globally-averaged ocean depth is ~ 3 km. Thus, the above estimation demonstrates that if such a steam atmosphere existed together with a comparable amount of H$_2$ on the Earth at the time of formation, then the current D/H ratio in the ocean must be higher by at least 30−70% than its initial value.

## 5. Evolution of D/H Ratios

As described in section 3, in reality, deuterium exchange between the ocean and the atmosphere occurs after (and possibly during) the ocean formation. This exchange results in a further increase in the D/H ratio of the ocean, relative to that calculated in Section 4. For example, when the surface temperature decreases to 300K and the ocean isotopically equilibrates with the atmosphere, the D/H ratio in the ocean becomes higher by approximately 300% compared to its initial value (see Fig. 2a). Isotopic fractionation due to atmospheric escape also increases the D/H ratio in the ocean more. In this section we show how much the D/H ratio increases during atmospheric escape.

---

[1] The temperature just after the ocean is fully formed might be higher than 300K. However, the resultant D/H ratio in liquid water is not so different, because the value of $f_{w(l)}$ of the fully-formed ocean is determined by equilibrium values for higher temperatures.

A possible escape mechanism of the hydrogen-rich atmosphere is the hydrodynamic escape induced by intense solar XUV irradiation (e.g., Zahnle 1993). Observations of young Sun-like stars imply that the young Sun emitted ~ 100 times higher XUV than the present Sun (e.g., Zahnle and Walker 1982; Pepin 1991). Hydrogen molecules absorb that energy at high altitudes in the atmosphere, resulting in enormous hydrodynamic escape of the atmospheric hydrogen.

Because there is much hydrogen in the atmosphere, the atmospheric escape is energy-limited. The energy-limited escape flux of hydrogen molecules ($F_{H2}$) is obtained by balancing the absorbed UV energy with the energy needed to lift the escaping atmosphere out of the Earth's potential well (Watson *et al.* 1981);

$$F_{H_2}(t) = \frac{\varepsilon \, f_{UV}(t) \, R_p}{m_{H_2} G M_p}, \qquad (18)$$

where $\varepsilon$ is the escape efficiency, which means how large a proportion of received UV energy is available for escaping hydrogen molecules, $f_{UV}$ is the UV energy flux, $m_{H2}$ is the mass of a hydrogen molecule ($3.3 \times 10^{-27}$ kg), $G$ is the gravitational constant ($6.67 \times 10^{-11}$ N m$^2$ kg$^{-2}$), $M_p$ is the Earth's mass ($6.0 \times 10^{24}$ kg), and $R_p$ is the Earth's radius ($6.4 \times 10^6$ m). Although the value of $\varepsilon$ has been estimated by several researchers for some atmospheric models (e.g., 0.15–0.3 in Watson *et al.* 1981; 0.42 in Sekiya *et al.* 1980; 0.15–0.5 in Yelle 2004), it is still quite uncertain. Thus, we treat $\varepsilon$ as a parameter in this paper.

The UV energy flux is given as a function of time based on observations of Sun-like stars of various ages. Here we use the following expression given by Guinan and Ribas (2002) and Baraffe *et al.* (2004);

$$f_{UV}(t) = f_{XUV}(t) + f_\alpha(t), \qquad (19)$$

$$f_{XUV}(t) = \begin{cases} 6.13 \, t^{-1.19} f_{XUV\_p} & \text{if } t \geq 0.1 \text{ Gyr}, \\ f_{XUV}(0.1) & \text{if } t < 0.1 \text{ Gyr}, \end{cases} \qquad (20)$$

$$f_\alpha(t) = \begin{cases} 3.17 \, t^{-0.75} f_{\alpha\_p} & \text{if } t \geq 0.1 \text{ Gyr}, \\ f_\alpha(0.1) & \text{if } t < 0.1 \text{ Gyr}, \end{cases} \qquad (21)$$

where $t$ is the stellar age in Gyr, and $f_{XUV\_p}$ and $f_{\alpha\_p}$ are the present solar XUV (< 100 nm) and Lyman-α surface-averaged fluxes, respectively. For $f_{XUV\_p}$ and $f_{\alpha\_p}$, we adopt $8.5\times10^{-4}$ W m$^{-2}$ and $1.42\times10^{-3}$ W m$^{-2}$, respectively, which correspond to the fluxes at 1 AU (Woods and Rottman 2002). Equations (20) and (21) reproduce the present fluxes for $t$ = 4.5 Gyr. Using Eqs. (18) and (19), we calculate temporal change of the mass of the hydrogen atmosphere. The temporal changes for different two values of $\varepsilon$ are shown in the lower panels of Fig. 5 for the case of the initial atmospheric mass of $1.6\times10^{21}$ kg which corresponds to the mass 10 times as large as the total mass of hydrogen atoms contained in the present Earth's ocean. The atmospheric mass is shown to substantially decrease in the early stage because of large UV energy flux.

The escaping H$_2$ drags HD upward. The escape flux of HD ($F_{HD}$) for an isothermal structure of the atmosphere at high altitudes is approximately given by (Hunten et al. 1987)

$$F_{HD}(t) = \begin{cases} F_{H_2}(t) \times \dfrac{X_{HD}}{X_{H_2}} \times \left( \dfrac{m_c - m_{HD}}{m_c - m_{H_2}} \right) & \text{if } m_c > m_{HD}, \\ 0 & \text{if } m_c \leq m_{HD}, \end{cases} \qquad (22)$$

where $X_{HD}$ and $X_{H_2}$ are respectively the mole fractions of HD and H$_2$ in the atmosphere, $m_{HD}$ is the mass of an HD molecule, and $m_c$ is the so-called crossover mass that is the mass of the heaviest species that can be dragged to space by the escaping H$_2$ molecules. The crossover mass ($m_c$) is defined as (Hunten et al. 1987)

$$m_c = m_{H_2} + \frac{F_{H_2} kT}{g\, b(H_2, HD)}, \qquad (23)$$

where $k$ is the Boltzmann constant ($1.38\times10^{-23}$ J K$^{-1}$), $g$ is the gravitational acceleration (9.8 m s$^{-2}$), and $b(H_2, HD)$ is the binary diffusion coefficient between H$_2$ and HD. We use the form of $b(H_2, HD) = 4.48\times10^{19} T^{0.75}$ m$^{-1}$ s$^{-1}$, which is derived based on experimental data of $b(H_2, D_2)$ (see Appendix C). In principle, the D/H ratio of the atmosphere increases during the hydrogen escape, because lighter molecules preferentially escape over heavier molecules. The degree of the enrichment in D depends on $F_{H_2}$. When $F_{H_2}$ is extremely large (i.e., $m_c \gg m_{HD}$), $F_{HD}/X_{HD} = F_{H_2}/X_{H_2}$ from Eq. (22). This means that the D/H ratio changes minimally during the hydrodynamic escape for this case, that is, the mass fractionation between HD and H$_2$ does not occur. On the other hand, a smaller $F_{H_2}$ results in a larger mass fractionation and thus, the D/H ratio of the atmosphere increases further.

Now we define the deuterium enrichment factors of the hydrogen atmosphere ($f_A$) and the ocean ($f_O$) as

$$f_A(t) = f_h = \frac{(D/H)_h}{(D/H)_0} \quad \text{and} \quad f_O(t) = f_{w(l)} = \frac{(D/H)_{w(l)}}{(D/H)_0}, \qquad (24)$$

where $(D/H)_0$ is the initial D/H ratio of the system. Below we show the evolution of $f_O$

and $f_A$. In the following simulations, we assume that the ocean has already been fully formed and always equilibrates isotopically with the atmosphere during the escape, which are good approximations as described in Section 3.

The upper panels of Fig. 5 show the temporal changes of $f_O$ and $f_A$ for $\varepsilon = 0.13$ (a) and 0.25 (b). In both calculations in Fig. 5a and 5b, the temperature is assumed to be 300K, and the initial masses of the hydrogen atmosphere and the ocean are respectively $1.6 \times 10^{21}$ kg and $1.4 \times 10^{21}$ kg ($\sim M_{PO}$: the present Earth's ocean mass), which correspond to the initial molar $H_2/H_2O = 10$. Since the ocean is always in isotopic equilibrium with the atmosphere, $f_O$ increases together with $f_A$ whose increase is caused by the mass fractionation due to the hydrodynamic escape. The fractionations are terminated when the hydrogen atmosphere is completely lost at $\sim 3.5$ Gyr for $\varepsilon = 0.13$ and $\sim 0.5$ Gyr for $\varepsilon = 0.25$. As described above, the effect of the mass fractionation is smaller for larger $F_{H2}$, and correspondingly larger $\varepsilon$. Thus, the final value of $f_O$ is $\sim 8$ in the low-$\varepsilon$ case (Fig. 5a), while $\sim 4$ in the high-$\varepsilon$ case (Fig. 5b).

Figure 6 shows $f_O$ at $t = 4.5$ Gyr as a function of the escape efficiency $\varepsilon$ for different choices of the initial molar $H_2/H_2O$ ratio ($[H_2]/[H_2O])_{init}$ and the initial ocean mass $M_{O,init}$. The value of $f_O$ increases with decreasing $\varepsilon$ for the reason explained above. Each curve ends off at a value of $\varepsilon$, because beyond the value a large amount of hydrogen still survives at 4.5 Gyr, which is unrealistic. Figure 6 also shows that smaller $([H_2]/[H_2O])_{init}$ results in smaller $f_O$ because the initial amounts of deuterium in the atmosphere and ocean are smaller. Note that if we assume no HD escape, we can readily calculate the maximum value of $f_O$ from the value of $([H_2]/[H_2O])_{init}$. For example, in the cases that $([H_2]/[H_2O])_{init} = 10$ and 3, the maximum values are 11 and 4, respectively. Because of the escape of HD, the values of $f_O$ shown in Fig. 6 are always lower than the maximum ones.

Figure 6 also shows the dependence of $f_O$ on $M_{o,init}$ in the case of $([H_2]/[H_2O])_{init} = 10$; the dashed curve represents the case of $M_{o,init} = 0.5\ M_{PO}$. As seen in Fig. 6, larger $M_{o,init}$ results in larger $f_O$ for a given value of $\varepsilon$. For example, when $\varepsilon = 0.2$, $f_O$ are 5 and 3.5 for the cases of 1 $M_{PO}$ and 0.5 $M_{PO}$, respectively. The atmospheric loss takes longer time for the case of larger $M_{o,init}$. Since the solar-UV flux decreases with time (see Eqs. (19)–(21)), the atmosphere with large $M_{o,init}$ experiences slow escape, resulting in more significant mass fractionation. On the other hand, the largest value of $f_O$ ($\sim 9.5$) for $M_{o,init} = 0.5M_{PO}$ is found to be larger than that ($\sim 8$) for $M_{o,init} = 1M_{PO}$. In both cases the atmospheric loss takes 4.5 Gyr. However, $\varepsilon$ for the former case ($\sim 0.06$) is smaller than that for the latter case ($\sim 0.1$). This means in the former case the escape flux ($F_{H2}$) is smaller, so that the mass fractionation is more significant.

We have so far considered escapes of hydrogen in the form of molecules, that is, $H_2$ and HD. However, it is possible that hydrogen escapes in the form of atoms (H and

D) because of dissociation of $H_2$ and HD by UV radiation. Using the binary diffusion coefficient for D in H, that is, $b(H, D) = 7.183 \times 10^{19} \, T^{0.728}$ m$^{-1}$ s$^{-1}$ (see Appendix C), we calculate the isotopic fractionation in the case of escape in the form of atoms (see dotted curve in Fig. 6). The initial condition is the same as that for the solid curve labeled "10". Figure 6 illustrates that resultant $f_O$ for the H-D escape differs little from that for the $H_2$-HD escape. This ensures that even if the escaping gas is composed of mixture of molecules and atoms, the result does not change so much.

### 6. Discussion and Conclusions

We discuss the origin of the ocean on the Earth and the formation of the terrestrial planets in the solar system, from the viewpoint of the D/H ratio in water. Because the average D/H ratio in carbonaceous chondrites (CC's) is very close to that in the Earth's ocean, the CC's origin of water has been widely accepted so far (see Fig. 1). This paper has, however, demonstrated that it is crucial whether the Earth had a hydrogen-rich atmosphere, and showed that the apparent concordance does not necessarily mean that the seawater came from CC's.

We have found that if the Earth had initially an atmosphere composed of $H_2$ and $H_2O$ with molar ratio $H_2/H_2O > \sim 1$, the D/H ratio in the ocean increased through the following processes. The equilibrium constant of the reaction of deuterium exchange between hydrogen gas and water vapor indicates that the fraction of deuterium in water increases with decreasing temperature (see Section 2). While water vapor condenses significantly at relatively high temperatures, the resultant D/H ratio in liquid water is, at least, 1.3–1.7 times higher than the initial D/H ratio in the system (i.e., $f_O \sim 1.3–1.7$; see Section 4). This enrichment is possible lower limits, because low-temperature exchange continues after ocean formation. The ocean equilibrates isotopically with the hydrogen atmosphere quickly (i.e., in at most $\sim 10^5$ years; see Section 3); $f_O$ increases to $\sim 2–3$, the equilibrium value at 300 K (see Section 5). Furthermore, $f_O$ increases to 2–9 through mass fractionation during hydrodynamic escape of the atmosphere, depending on the escape efficiency (see Section 5).

If the atmospheric escape occurred relatively slowly (> 1Gyr), the scenario described above is consistent with recent theories with nebular gas for terrestrial planet formation in the solar system. Planets form through oligarchic growth, followed by runaway growth. After each protoplanet captures all the planetesimals in its feeding zone, about 20 protoplanets in circular orbits are formed in the terrestrial planet region (Kokubo and Ida 1998, 2000). Subsequent growth of protoplanets requires giant collisions between those protoplanets, resulting in eccentric planets (e.g., Kokubo *et al.* 2006; Chambers *et al.* 1996). The high eccentricities are damped by the drag of the nebular gas (Kominami and Ida 2002; Nagasawa *et al.* 2005). After the final giant collision the Earth attracts the nebular gas gravitationally to have a massive hydrogen-rich atmosphere of more than $10^{21}$ kg (Hayashi *et al*. 1979; Ikoma and Genda 2006). Then, the atmospheric hydrogen is oxidized by some oxides such as FeO in the magma ocean to produce water (Ikoma and Genda 2006). The molar $H_2/H_2O$ ratio in the atmosphere is ~ 10 (e.g., Kuramoto and Matsui 1996). As described above, such an initial state results in the D/H ratio in the ocean 2–9 times higher than the initial D/H ratio, while the D/H ratio of the nebular gas is known to be lower by a factor of 6–9 than that of the ocean on the Earth (see Fig. 1).

In contrast, nebular-gas-free accretion theories (e.g., Raymond *et al*. 2004, 2005) seem to be inconsistent with the current isotopic composition of the ocean on the Earth. In those theories, CC-like planetesimals are delivered to the formation site of the Earth because of gravitational perturbation of Jupiter. The planetesimals deliver water and damp the eccentricities of planets via dynamical friction. As described in Section 1, recent calculations of chemical-equilibrium compositions of gas from CC's (Hashimoto *et al*. 2007) demonstrate that a hydrogen-rich atmosphere is likely to form on the Earth even in that situation. As shown in Section 4, the D/H ratio in the ocean increases, at least, by a factor of 1.3–1.7, while the D/H ratio of CC's is equal to that in the ocean within 10%.

While the results obtained in this paper seem to support the nebular origin of water on the Earth, only those are insufficient for us to reach the conclusion and to exclude the chondritic origin of water. A crucial uncertainty is the escape efficiency. That depends on both the UV flux and the atmospheric composition through the complex chemical reactions and dynamics (e.g., Yelle 2004). Thus, we have treated it as a parameter. Another uncertainty is how much water dissolved into the magma ocean and what fraction of the water returned to the surface. Water is known to dissolve into molten silicate at high temperatures and pressures (e.g., Fricker and Reynolds 1968). As shown in Fig. 2a, water condensed at high temperatures preserves its pristine value. Thus, if much water compared to atmospheric water dissolved into the magma ocean and the water returned to the surface later, then the current D/H ratio in the ocean is hard to be explained only in the context of the nebular hypothesis. On the other hand, dynamical modeling of planet formation that takes into account contents of iron, iron oxides, and organics would be needed to check whether a hydrogen-rich atmosphere is really formed. Geochemical studies are also needed. In this respect, the D/H ratios in organic matter in the sediments deposited over the last 3.5 Gyr are reported to be similar to that in the present Earth's ocean within 10 %, but whether they have suffered contamination of the current seawater is yet uncertain (Robert 1989).

# APPENDIX

## Appendix A  Characteristic reaction timescale for isotopic equilibrium

The temporal change of the partial pressure of each species due to reaction (1) is given by

$$\frac{dp_{HDO}}{dt} = \frac{dp_{H_2}}{dt} = -\frac{dp_{HD}}{dt} = -\frac{dp_{H_2O}}{dt} = k^+ p_{HD} p_{H_2O} - k^- p_{H_2} p_{HDO}. \tag{A1}$$

Differentiating Eq. (3) with respect to time, we obtain

$$\frac{d}{dt}\left(\frac{D}{H}\right)_{w(g)} = \frac{1}{2p_{H_2O} + p_{HDO}}\left\{\frac{dp_{HDO}}{dt} - \left(\frac{D}{H}\right)_{w(g)}\left(2\times\frac{dp_{H_2O}}{dt} + \frac{dp_{HDO}}{dt}\right)\right\}. \tag{A2}$$

Using Eq. (A1), we can eliminate the time derivatives of the partial pressures in Eq. (A2);

$$\frac{d}{dt}\left(\frac{D}{H}\right)_{w(g)} = p_{H_2}\left\{1+\left(\frac{D}{H}\right)_{w(g)}\right\}\left\{k^+ \times \frac{p_{H_2O}}{2p_{H_2O}+p_{HDO}} \times \frac{p_{HD}}{p_{H_2}} - k^- \times \left(\frac{D}{H}\right)_{w(g)}\right\}. \tag{A3}$$

From Eq. (3),

$$p_{HDO} = \frac{2\times(D/H)_{w(g)}}{1-(D/H)_{w(g)}} \times p_{H_2O}, \quad \text{and} \quad p_{HD} = \frac{2\times(D/H)_h}{1-(D/H)_h} \times p_{H_2}. \tag{A4}$$

Inserting Eq. (A4) into Eq. (A3), we finally obtain

$$\frac{d}{dt}\left(\frac{D}{H}\right)_{w(g)} = p_{H_2}\left\{1+\left(\frac{D}{H}\right)_{w(g)}\right\}\left\{k^+ \times \frac{1-(D/H)_{w(g)}}{1-(D/H)_h} \times \left(\frac{D}{H}\right)_h - k^- \times \left(\frac{D}{H}\right)_{w(g)}\right\}. \tag{A5}$$

In the same way, we obtain

$$\frac{d}{dt}\left(\frac{D}{H}\right)_h = -p_{H_2O}\left\{1+\left(\frac{D}{H}\right)_h\right\}\left\{k^+ \times \left(\frac{D}{H}\right)_h - k^- \times \frac{1-(D/H)_h}{1-(D/H)_{w(g)}} \times \left(\frac{D}{H}\right)_{w(g)}\right\}. \tag{A6}$$

Since we assume $(D/H)_{w(g)} \ll 1$ and $(D/H)_h \ll 1$, Eqs. (A5) and (A6) are simplified as

$$\begin{cases} \dfrac{d}{dt}\left(\dfrac{D}{H}\right)_{w(g)} = p_{H_2}\left\{k^+ \times \left(\dfrac{D}{H}\right)_h - k^- \times \left(\dfrac{D}{H}\right)_{w(g)}\right\}, \\ \dfrac{d}{dt}\left(\dfrac{D}{H}\right)_h = -p_{H_2O}\left\{k^+ \times \left(\dfrac{D}{H}\right)_h - k^- \times \left(\dfrac{D}{H}\right)_{w(g)}\right\}. \end{cases} \tag{A7}$$

We define the difference from the D/H ratio at the isotopic equilibrium, $(D/H)^d$, as follows;

$$\left(\frac{D}{H}\right)_{w(g)} = \left(\frac{D}{H}\right)^e_{w(g)} + \left(\frac{D}{H}\right)^d_{w(g)} \quad \text{and} \quad \left(\frac{D}{H}\right)_h = \left(\frac{D}{H}\right)^e_h + \left(\frac{D}{H}\right)^d_h. \tag{A8}$$

where $(D/H)^e$ is the D/H ratio at the isotopic equilibrium. Inserting Eq. (A8) into Eq. (A7), we obtain

$$\begin{cases} \dfrac{d}{dt}\left(\dfrac{D}{H}\right)^d_{w(g)} = p_{H_2}\left\{k^+ \times \left(\dfrac{D}{H}\right)^d_h - k^- \times \left(\dfrac{D}{H}\right)^d_{w(g)}\right\}, \\ \dfrac{d}{dt}\left(\dfrac{D}{H}\right)^d_h = -p_{H_2O}\left\{k^+ \times \left(\dfrac{D}{H}\right)^d_h - k^- \times \left(\dfrac{D}{H}\right)^d_{w(g)}\right\}. \end{cases} \quad (A9)$$

From the conservation of deuterium molecules, we obtain the following relation:

$$p^e_{HD} + p^e_{HDO} = p_{HD} + p_{HDO}, \quad (A10)$$

where $p^e_X$ is the isotopically equilibrated partial pressure of species X. Inserting Eq. (A4) into Eq. (A10) on the assumption of $(D/H)_{w(g)} \ll 1$ and $(D/H)_h \ll 1$, we obtain

$$\left(\dfrac{D}{H}\right)_h p_{H_2} + \left(\dfrac{D}{H}\right)_{w(g)} p_{H_2O} = \left(\dfrac{D}{H}\right)^e_h p^e_{H_2} + \left(\dfrac{D}{H}\right)^e_{w(g)} p^e_{H_2O}. \quad (A11)$$

Since $p^e_{H2} = p_{H2}$ and $p^e_{H2O} = p_{H2O}$ on the assumption of $(D/H)_{w(g)} \ll 1$ and $(D/H)_h \ll 1$,

$$\left(\dfrac{D}{H}\right)^d_{w(g)} p_{H_2O} = -\left(\dfrac{D}{H}\right)^d_h p_{H_2}. \quad (A12)$$

Eqs. (A9) are rewritten as

$$\begin{cases} \dfrac{d}{dt}\left(\dfrac{D}{H}\right)^d_{w(g)} = -\left(k^+ p_{H_2O} + k^- p_{H_2}\right) \times \left(\dfrac{D}{H}\right)^d_{w(g)}, \\ \dfrac{d}{dt}\left(\dfrac{D}{H}\right)^d_h = -\left(k^+ p_{H_2O} + k^- p_{H_2}\right) \times \left(\dfrac{D}{H}\right)^d_h. \end{cases} \quad (A13)$$

Inserting Eq. (A13) into Eq. (6), we obtain Eq. (7).

## Appendix B   Derivation of Eqs. (13)–(16)

We consider a box with volume $V$ containing hydrogen molecules and water molecules. The total masses of hydrogen gas ($M_h$) and water vapor ($M_{w(g)}$) are, respectively, given by

$$M_h = V\rho_h \quad \text{and} \quad M_{w(g)} = V\rho_{w(g)}(T), \quad (B1)$$

where $\rho_h$ and $\rho_{w(g)}$ are the partial densities of hydrogen gas and water vapor, respectively. Since hydrogen molecules are not condensed in the temperature range considered here, $\rho_h$ is constant with temperature. Since we have neglected the volume of liquid water (assumption (1) in Section 4), the total mass of liquid water ($M_{w(l)}$) is given by

$$M_{w(l)} = V\{\rho_{w(g)}(T_0) - \rho_{w(g)}(T)\}, \quad (B2)$$

where $T_0$ is the temperature at which the water vapor starts to condense.

We consider conservation of the number of deuterium molecules ($N_D$). Below we assume $p_{H2} \gg p_{HD}$ and $p_{H2O} \gg p_{HDO}$. Since there are no liquid water molecules at the temperature $T_0$, $N_D(T_0)$ is given by

$$N_\mathrm{D}(T_0) = V \times \left[ \frac{\rho_\mathrm{h}}{m_\mathrm{H}} + \frac{2\rho_\mathrm{w(g)}(T_0)}{18 m_\mathrm{H}} \right] \times (\mathrm{D/H})_0, \tag{B3}$$

where $m_\mathrm{H}$ is the mass of a hydrogen atom, and $(\mathrm{D/H})_0$ is the D/H in the box. The number of deuterium molecules at $T$ ($< T_0$) is given by

$$N_\mathrm{D}(T) = \frac{2}{18} \frac{M_\mathrm{w(l)}}{m_\mathrm{H}} \times (\mathrm{D/H})_\mathrm{w(l)} + \frac{2}{18} \frac{M_\mathrm{w(g)}}{m_\mathrm{H}} \times (\mathrm{D/H})_\mathrm{w(g)} + \frac{V \rho_\mathrm{h}}{m_\mathrm{H}} (\mathrm{D/H})_\mathrm{h}, \tag{B4}$$

where $(\mathrm{D/H})_\mathrm{w(l)}$, $(\mathrm{D/H})_\mathrm{w(g)}$, and $(\mathrm{D/H})_\mathrm{h}$ are the D/H ratios of liquid water, water vapor, and hydrogen gas, respectively. From $N_\mathrm{D}(T) = N_\mathrm{D}(T_0)$, the following equation is obtained,

$$\rho_\mathrm{h} + \frac{\rho_\mathrm{w(g)}(T_0)}{9} = f_\mathrm{h} \rho_\mathrm{h} + \frac{f_\mathrm{w(l)}}{9} \{\rho_\mathrm{w(g)}(T_0) - \rho_\mathrm{w(g)}(T)\} + \frac{f_\mathrm{w(g)}}{9} \rho_\mathrm{w(g)}(T), \tag{B5}$$

where $f_X$ is the deuterium enrichment factor of the species X, which is defined by Eq. (12).

Since we have assumed the isotopic equilibrium between hydrogen molecules and water vapor molecules (assumption (2) in Section 4), Eq. (4) can be used. Using $f$, we rewrite Eq. (4) as follows;

$$f_\mathrm{w(g)} = K f_\mathrm{h}. \tag{B6}$$

Next we derive the equation for $f_\mathrm{w(l)}$. Since we have assumed that the D/H ratio of water molecules condensed at a given temperature has the D/H ratio of water vapor molecules at that temperature and the D/H ratio of bulk liquid water changes only by addition of newly condensed water molecules (assumption (3) in Section 4), a very small change of $N_\mathrm{D\text{-}w(l)}$ (the number of deuterium molecules in liquid water) is given by

$$dN_\mathrm{D\text{-}w(l)} = (\mathrm{D/H})_\mathrm{w(g)} \frac{2}{18 m_\mathrm{H}} \times (-dM_\mathrm{w(g)}), \tag{B7}$$

where $dM_\mathrm{w(g)}$ is a very small change of the mass of water vapor ($M_\mathrm{w(g)}$). Since $dM_\mathrm{w(g)} = -dM_\mathrm{w(l)}$, where $dM_\mathrm{w(l)}$ is a very small change of the mass of liquid water,

$$\frac{dN_\mathrm{D\text{-}w(l)}}{dT} = (\mathrm{D/H})_\mathrm{w(g)} \frac{2}{18 m_\mathrm{H}} \times \frac{dM_\mathrm{w(l)}}{dT}. \tag{B8}$$

The number of deuterium molecules in liquid water ($N_\mathrm{D\text{-}w(l)}$) is given by

$$N_\mathrm{D\text{-}w(l)} = (\mathrm{D/H})_\mathrm{w(l)} \times \frac{2 M_\mathrm{w(l)}}{18 m_\mathrm{H}}. \tag{B9}$$

From Eqs. (B8), (B9), and (12),

$$\frac{df_\mathrm{w(l)}}{dT} = (f_\mathrm{w(g)} - f_\mathrm{w(l)}) \frac{1}{M_\mathrm{w(l)}} \frac{dM_\mathrm{w(l)}}{dT}. \tag{B10}$$

**Appendix C  Estimation of the binary diffusion coefficients**

The binary diffusion coefficient for HD in $H_2$ has not been measured directly but can be calculated from the known value for $D_2$ in $H_2$. According to Mason and Marrero (1970), $b(H_2, D_2)$ is given by

$$b(H_2, D_2) = 1.81 \times 10^{22} \, T^{0.500} \times \{\ln(6.36 \times 10^6/T)\}^{-2} \times \exp(-6.072/T - 38.10/T^2) \, \text{m}^{-1} \, \text{s}^{-1}, \quad (C1)$$

where $T$ is the temperature in Kelvin. For simplicity, we approximate Eq. (C1) as the formulation of $b = AT^s$, and then obtain as

$$b(H_2, D_2) = 4.25 \times 10^{19} \, T^{0.75} \, \text{m}^{-1} \, \text{s}^{-1}. \quad (C2)$$

The relative error between Eq. (C1) and Eq. (C2) is within 3% for the range of 200–2000K. The diffusion coefficient for HD in $H_2$ should be larger than $b(H_2, D_2)$ by a factor of the square root of the reduced mass (e.g., Kasting and Pollack 1983), that is,

$$\frac{b(H_2, HD)}{b(H_2, D_2)} = \sqrt{\left(\frac{m_{H_2} + m_{HD}}{m_{H_2} m_{HD}}\right) \bigg/ \left(\frac{m_{H_2} + m_{D_2}}{m_{H_2} m_{D_2}}\right)} \cong 1.054. \quad (C3)$$

where $m_X$ is the mass of molecule, and the subscript X represents each particular species. We approximate $b(H_2, HD)$ by

$$b(H_2, HD) = 4.48 \times 10^{19} \, T^{0.75} \, \text{m}^{-1} \, \text{s}^{-1}. \quad (C4)$$

In addition to $b(H_2, HD)$, we use $b(H, D)$ in Section 5, which has also been not measured directly. In the same way, we can estimate $b(H, D)$ from $b(H, H)$ that has been given by Weissman and Masson (1962) as follows:

$$b(H, H) = 8.295 \times 10^{19} \, T^{0.728} \, \text{m}^{-1} \, \text{s}^{-1}. \quad (C5)$$

We can obtain

$$b(H, D) = 7.183 \times 10^{19} \, T^{0.728} \, \text{m}^{-1} \, \text{s}^{-1}. \quad (C6)$$

## ACKNOWLEDGEMENTS

We are grateful to S. Ida, M. Fujimoto, and H. Yurimoto for fruitful discussions and their continuous encouragement. We thank F. Robert for comments on the isotopic reaction rate. We also thank C. Parkinson and the other anonymous reviewers for comments and suggestions. This research was partly supported by the 21st Century COE Program "How to build habitable planets", Tokyo Institute of Technology, by Grand-in-Aid for Scientific Research on Priority Areas, both of which are sponsored by the Ministry of Education, Culture, Sports, Technology and Science (MEXT), Japan, and by the Research Fellowship of the Japan Society for the Promotion of Science for Young Scientists.

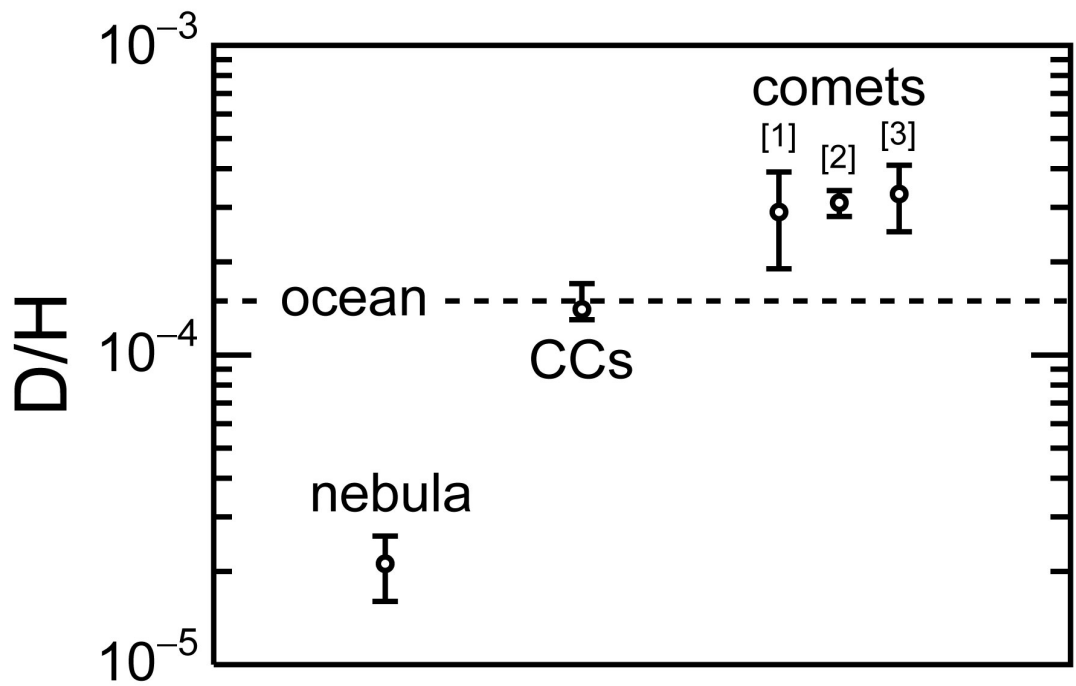

**Figure 1.** The D/H ratios in the present Earth's ocean (Lécuyer *et al*. 1998), protosolar H$_2$ [nebula] (Geiss and Gloeckler 1998), carbonaceous chondrites [CCs] (Robert *et al*. 2000), and three comets; [1] Hyakutake (Bockelée *et al*. 1998), [2] P/Halley (Eberhardt *et al*. 1995), [3] Hale-Bopp (Meier *et al*. 1998).

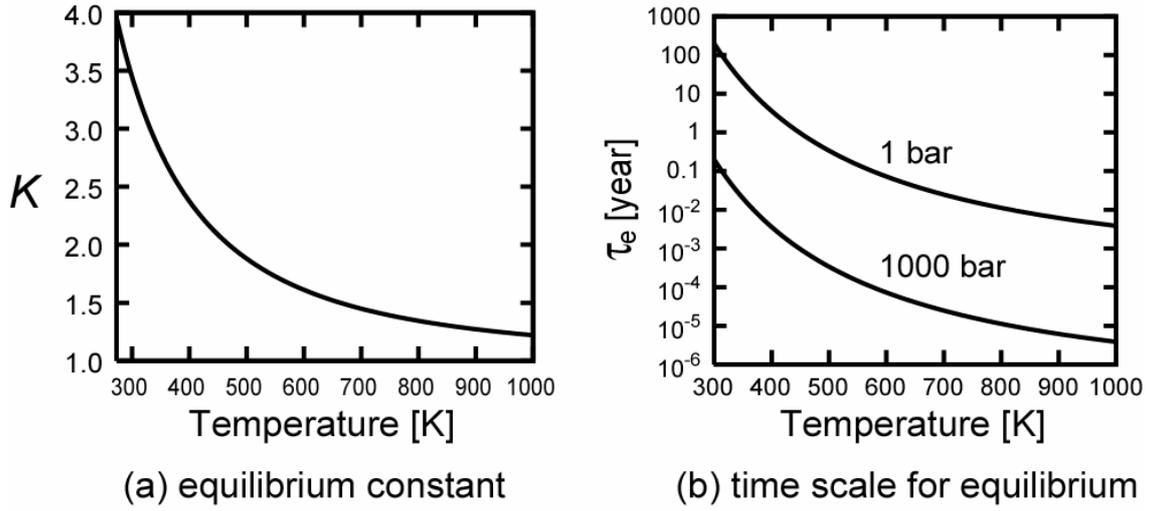

**Figure 2.** (a)The equilibrium constant for deuterium exchange reaction between hydrogen gas and water vapor given by Eq. (5), and (b) time scale for isotopic equilibrium between hydrogen gas and water vapor given by Eq. (8). The pressures in right panel represent the total pressures ($p_t$).

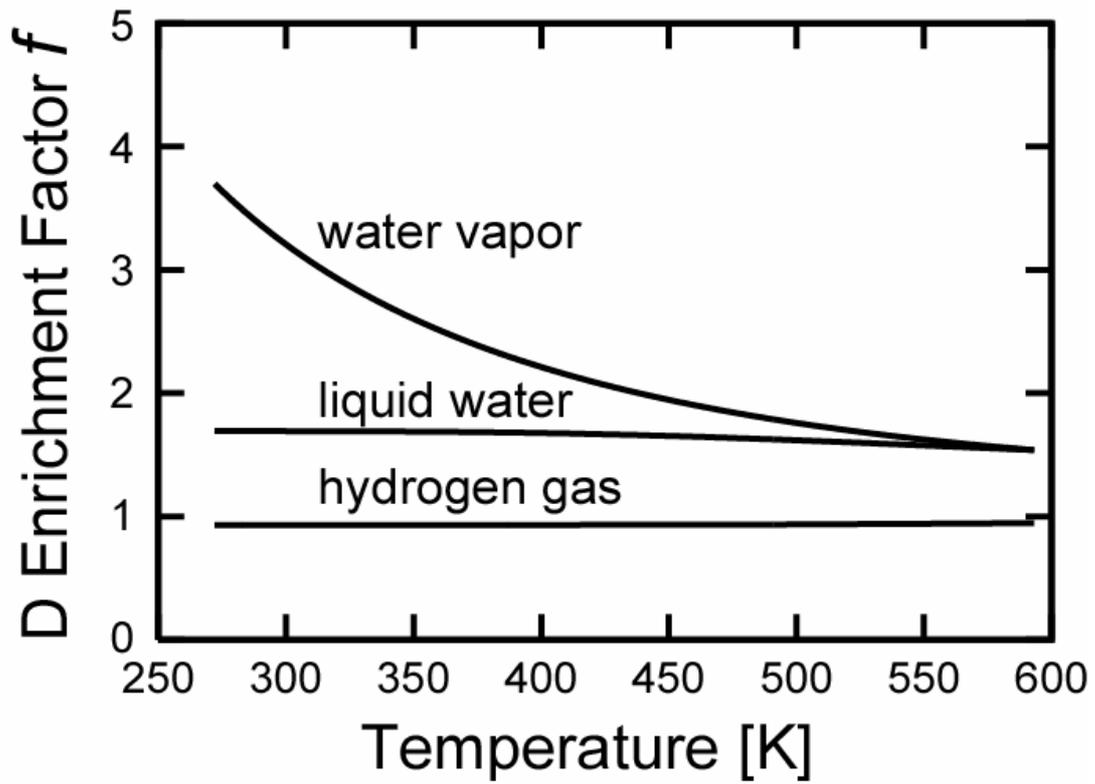

**Figure 3.** The changes in the deuterium enrichment factors of the liquid water ($f_{w(l)}$), the water vapor ($f_{w(g)}$), and the hydrogen gas ($f_h$) by cooling during the stage of ocean formation in the case of $p_{w(g)}(T_0)$ = 200 bar and molar $H_2/H_2O$ = 10.

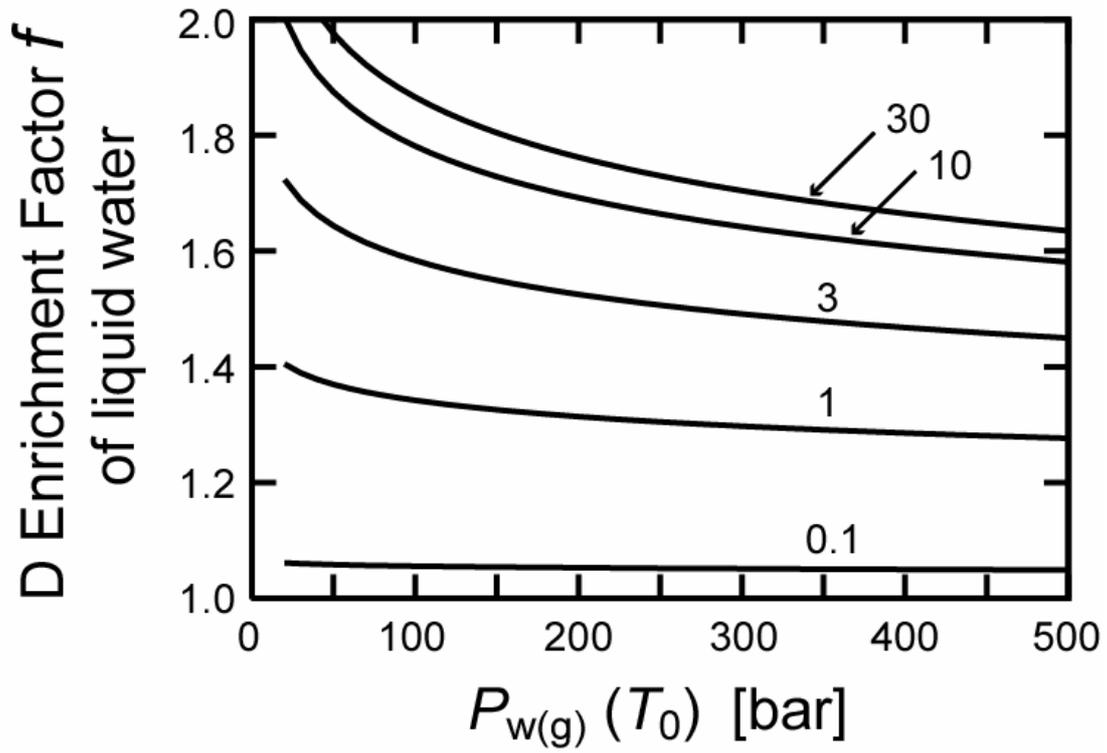

**Figure 4.** The deuterium enrichment factor of liquid water (ocean), $f_{w(l)}$, at 300 K for various values of the vapor pressure at $T_0$, $P_{w(g)}(T_0)$, and the molar $H_2/H_2O$ ratios. The value attached to each curve represents the molar $H_2/H_2O$ ratios. Note that the present Earth's ocean mass corresponds to ~ 300 bar.

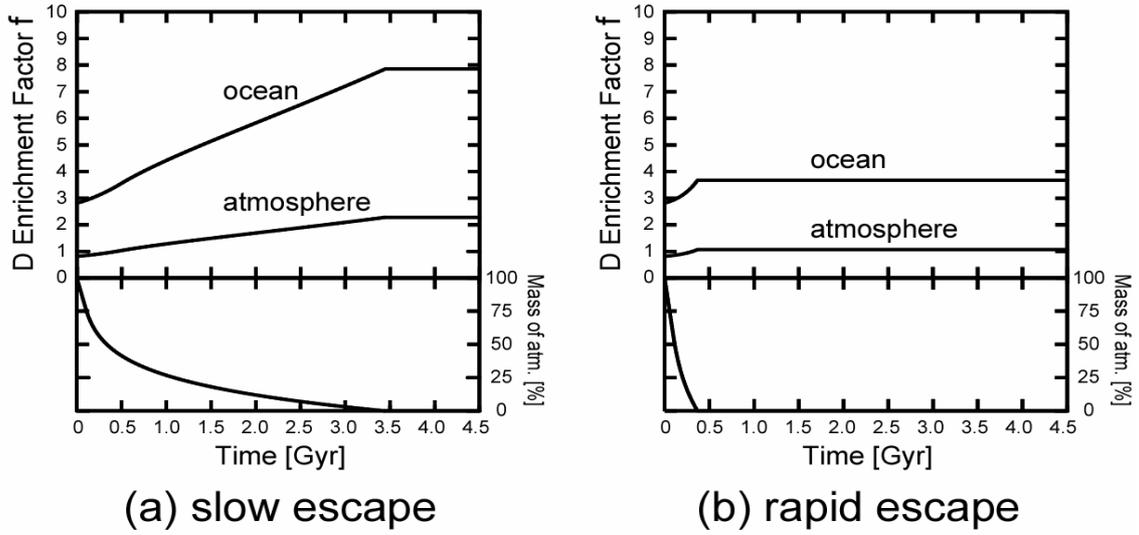

**Figure 5.** The evolution of the deuterium enrichment factors of the ocean ($f_O$) and the atmosphere ($f_A$) (upper panels), and the evolution of the mass of the hydrogen atmosphere during hydrodynamic escape of the hydrogen atmosphere (lower panels). In these calculations, the mass of the ocean is the present Earth's ocean mass ($1.4 \times 10^{21}$ kg), initial molar $H_2/H_2O$ ratio is 10, and the escape efficiency $\varepsilon$ is (a) 0.13 and (b) 0.25.

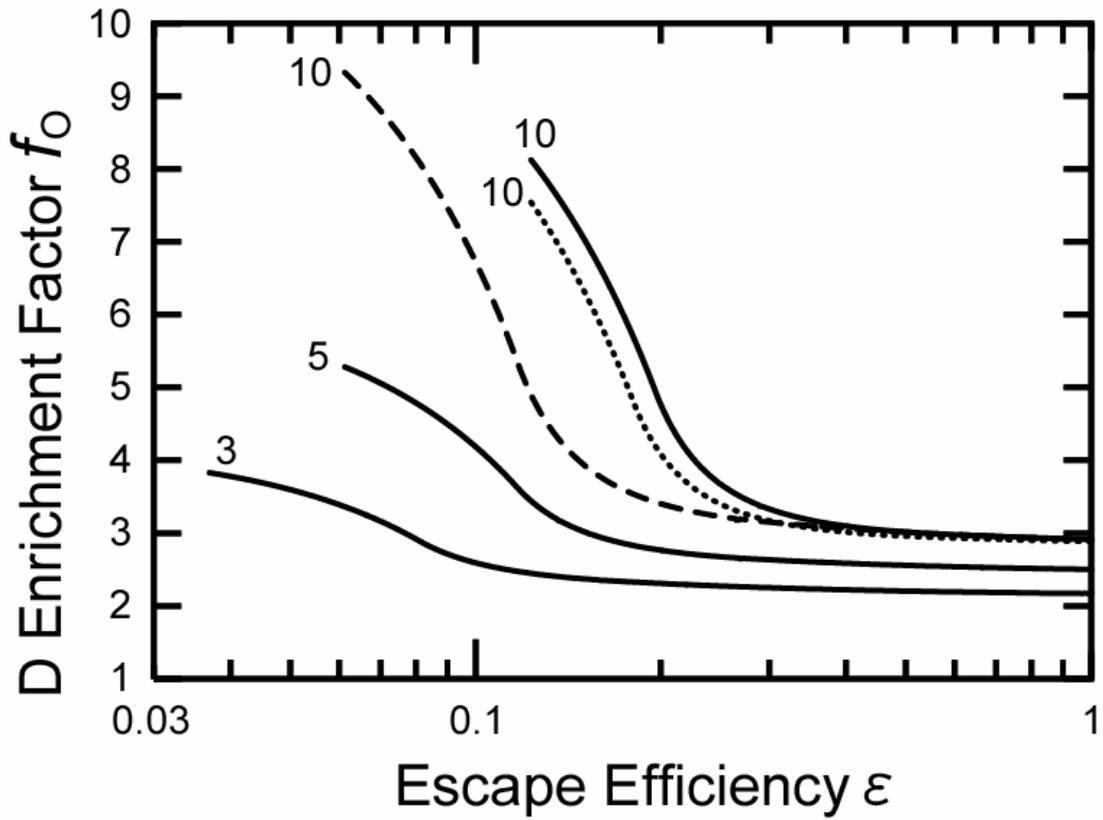

**Figure 6.** The deuterium enrichment factors of the ocean ($f_O$) after 4.5 Gyr as a function of escape efficiency $\varepsilon$. The value attached to each curve represents the initial molar $H_2/H_2O$ ratios. The mass of the present Earth's ocean ($M_{PO}$) is considered in the cases except for dashed curve. For dashed curve $0.5 M_{PO}$ is considered. The dotted curve represents the hydrodynamic escape in the form of atoms, that is, H and D.